\documentclass[twocolumn,showpacs,preprintnumbers,amsmath,amssymb]{revtex4}

  \usepackage{graphicx}
  \usepackage{amsmath}
  \usepackage{amsfonts}
  \usepackage{amssymb}
  \usepackage{nicefrac}

  \newcommand{\vecd}[1]{\mathbf{#1}}
  \newcommand{\vecbold}[1]{\boldsymbol{#1}}
  \newcommand{\tens}[1]{\sf {#1}}
  \newcommand{\mat}[1]{\tens{#1}}

\begin{document}

\title{Dumbbell diffusion in a spatially periodic potential}

\author{Jochen Bammert, Steffen Schreiber and Walter Zimmermann}

\affiliation{Theoretische Physik I, Universit\"at Bayreuth, D-95440 Bayreuth, Germany}

\date{September 1, 2008}

\begin{abstract}
We present a numerical investigation of the Brownian motion and diffusion of 
a dumbbell in a two-dimensional periodic potential. Its dynamics is described by a
Langevin model including the hydrodynamic interaction. With increasing values of the amplitude
of the potential we find along the modulated spatial
directions a reduction of the diffusion constant and of the impact of the hydrodynamic interaction.
For modulation amplitudes of the potential in the range of the thermal energy 
the dumbbell diffusion exhibits
a pronounced local maximum  at 
a wavelength of about 3/2 of the dumbbell extension.
This is especially emphasized for stiff springs connecting the two beads.
\end{abstract}


\maketitle

{\it{Introduction.-}}\label{sec: intro} Investigations on the diffusion 
of different colloidal particles in a homogeneous solvent have a long history \cite{Einstein:1905.1,DoiEd},
while studies on the diffusion of small spheres, dimers and polymers in different potentials attract considerable interest only for a short time \cite{Grier:2002.1,Dholakia:2003.1,Ruckenstein:1995,Grier:2004.1,Lacasta:2004.1,Lacasta:2005.1,Fasolino:2005}.
Laser-tweezer arrays are a new powerful tool for
generating the desired 
spatially periodic, correlated or unstructured potentials in order to study the effects 
of inhomogeneous potential landscapes on the motion of colloidal particles
\cite{Grier:2002.1,Dholakia:2003.1,Grier:2004.1,Grier:2007.1}. 
Furthermore recent studies of dumbbells and polymers in 
random potentials are exciting 
issues in statistical physics \cite{Trimper:2001.1,Vilgis:2004.1}.

Several of these investigations are motivated 
by possible applications like particle sorting in inhomogeneous potentials. 
For example cross-streamline migration of colloidal particles has been found
in a flow through an optically induced periodic potential. 
Since this migration 
depends on the extent of the colloidal particles, the laser-tweezer array
has been successfully used for sorting particles with respect to their size \cite{Dholakia:2003.1,Grier:2007.1}.

We investigate the Brownian motion and the diffusion of 
dumbbells through a two-dimensional periodic potential, which is described by a Langevin model. 
In doing so we include the effects
of the hydrodynamic interaction between the two beads of the dumbbell and focus on the interplay between the dumbbell extension $b$ and the
wavelength $\lambda$ of the spatially periodic potential. In the context of this
work the dumbbell may be considered as a simple model for several deformable objects such as pom-pom polymers \cite{McLeish:98.1} or two colloids which are connected either by a rather flexible $\lambda$-DNA molecule or by a semi-flexible actin filament.

Our numerical studies reveal a significant 
dependence of the dumbbell diffusion on the
ratio $\lambda/b$. With the potential amplitude $V_0$
of the order of the thermal excitation energy $k_BT$ 
we find a remarkable maximum of the dumbbell diffusion constant in the range 
of $\lambda \approx 3b/2$,  whereby the height of the diffusion maximum 
increases with the stiffness of the spring connecting
the two beads of the dumbbell. Another remarkable effect is the reduction of the influence of the hydrodynamic interaction with increasing potential amplitude.

\vspace{5mm}

{\it Model.-}\label{sec: numeric}
We describe the Brownian motion 
of a dumbbell in a two-dimensional periodic potential 
by the following Langevin equation (without inertia)
\begin{equation}
 \dot{\vecd{r}}_i = {\mat{H}}_{ij} \left( {\vecd{F}}^{\Phi}_j + \vecd{F}^V_j \right) +\vecd{F}^S_i\,,\qquad (i,j=1,2)
\end{equation}
for the bead positions  $\vecd{r}_i=(x_i,y_i,z_i)$. The linear spring force $\vecbold{F}^{\Phi}_i$ between them is determined by the harmonic potential
\vspace*{-4mm}
\begin{eqnarray}
 \Phi(\vecd{r}_1,\vecd{r}_2) = \frac{k}{2}\, \left(b-| \vecd{r}_1 -\vecd{r}_2|\right)^2\,,
\end{eqnarray}
with the  equilibrium length $b$
of the spring and the corresponding 
spring constant $k$. The spatially periodic
force $\vecd{F}_i^V=-\nabla V(\vecd{r}_i)$ is derived from the two-dimensional potential in the $xy$-plane
\begin{equation}
 \label{perpot}
 V(\vecd{r}_i) = 2V_0 \cos\left(\frac{x_i+y_i}{\lambda} \pi \right) \cos\left(\frac{x_i-y_i}{\lambda} \pi \right)\,,
\end{equation}
which can be realized in experiments by a laser tweezer 
array. Its amplitude $V_0$ may be changed by varying 
the intensity of the laser light. 
The same  wavelength $\lambda$ is chosen in the  $x$- and in the $y$-direction.

In the absence of hydrodynamic interaction ({\bf HI}) 
between the beads the mobility matrix $\mat{H}$ is a diagonal 
matrix ($\mat{H}_{ii}=\frac{1}{\zeta}\,\mat{I} \,, \mat{H}_{ij}=0$ \, for $i \neq j$) being inversely 
proportional to the Stokes friction coefficient $\zeta=6 \pi \eta a$ 
which depends on the solvent viscosity $\eta$ as well as on the effective hydrodynamic bead radius $a$.
The HI between the two beads is taken into account by the Rotne-Prager tensor \cite{RotnePrager:1969} where the mobility matrix for $i \neq j$ has the following structure:
\begin{align}
 \hspace{-1mm}{\mat{H}}_{ij}= \frac{1}{8\pi\eta r_{ij}} \left[ \left(1+\frac{2}{3}\frac{a^2}{r_{ij}^2} \right){\mat{I}} + \left( 1 - 2 \frac{a^2}{r_{ij}^2} \right) \hat{\vecd{r}}_{ij} \hat{\vecd{r}}_{ij}^T \right]   .
 \end{align}
Note that $\vecd{r}_{ij}=\vecd{r}_i-\vecd{r}_j$ is the distance vector between the beads and $r_{ij}$ is its norm.

The stochastic forces $\vecd{F}^S_i$ caused by the thermal heat bath are related to the dissipative 
drag by the fluctuation dissipation theorem 
which ensures the correct equilibrium properties. They can be combined to a single supervector $\vecd{F}^S=(\vecd{F}_1^S,\vecd{F}_2^S)$ that reads
\begin{eqnarray}
\vecd{F}^S= \sqrt{ 2 k_B T {\mat{H}}}\, \vecbold{\xi}\,.
\end{eqnarray}
$T$ is the temperature, $k_B$ the Boltzmann constant and 
${\vecbold{\xi}}(t)$ is the uncorrelated Gaussian white noise vector with zero mean and unit variance
\begin{eqnarray}
\left< \vecbold{\xi}(t)  \right>&=&0 \,, \\
\left< \vecbold{\xi}(t)\vecbold{\xi}^T(t') \right>&=&\delta(t-t')\,{\mat{I}}\,.
\end{eqnarray}
The fixed parameters in our simulations are $b=1$ for the typical length scale and $k_BT=1$ which determines the energy scale. If not stated otherwise we use $\frac{a}{b}=\frac{1}{5}$, $\eta=1$. $k$ determines the binding energy of the spring in units of the thermal energy $\frac{kb^2}{k_BT}$. Most of our results were observed after averaging over more than $10^4$ ensembles.

\begin{figure}[ht]
\vspace{-4mm}
  \begin{center}
  \includegraphics[width=0.9\columnwidth]{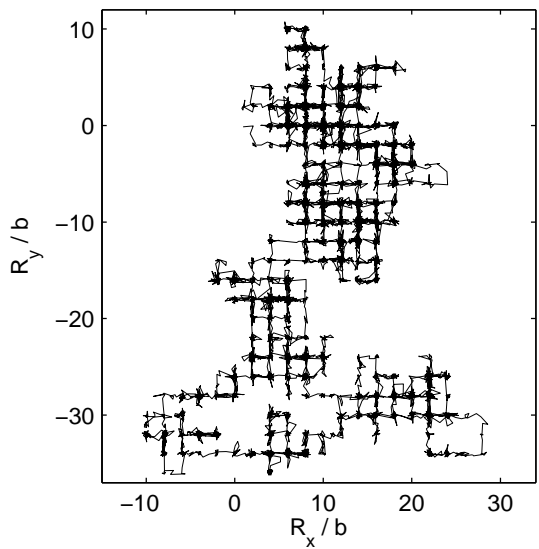}
  \end{center}
\vspace{-5mm}
  \caption{The trajectory of the CM of the dumbbell 
in the $xy$-plane follows predominantly along the saddles between
the minima of the potential.  Parameters: $k=10$, $V_0/k_BT=2$, and 
$\lambda=2b$.}
\label{traj}
\end{figure}

{\it Results.-}\label{sec: results}
A typical trajectory of the center of mass of the dumbbell ({\bf CM}), ${\bf R}=({\bf r}_1+{\bf r}_2)/2$,
in the $xy$-plane is shown in  Fig.~\ref{traj} for $k=10$, $\lambda=2b$ and $V_0=2k_BT$. It passes the saddles 
between the maxima of the potential and accordingly the 
trajectory adapts to the quadratic structure of the potential landscape.
If one of the two parameters
$k$ or  $\lambda / b$ is reduced, 
larger  excursions of the CM away from the potential minima
occur and even diagonal jumps between the valleys are found frequently.

\begin{figure}[ht]
\vspace{-2mm}
  \begin{center}
  \includegraphics[width=0.95\columnwidth]{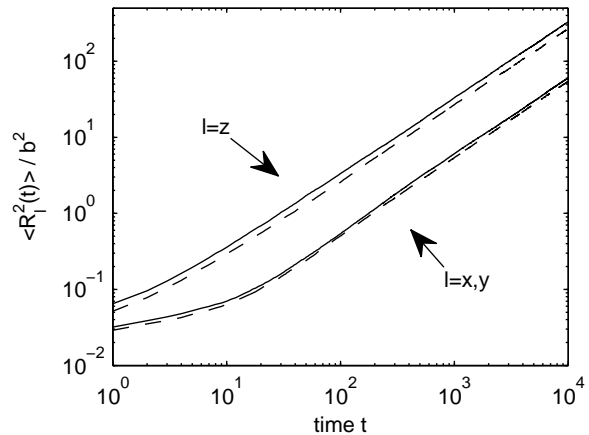}
  \end{center}
\vspace{-5mm}
  \caption{The mean square displacement of the CM of the dumbbell 
is shown along the $x$- and the $y$-direction (lower lines) 
 and along the $z$-direction (upper lines) with HI
between the beads  (solid lines) and without (dashed lines). 
The parameters are the same as in Fig.~\ref{traj}.}
\label{msquaredist}
\end{figure}

The mean square displacement $\langle R_l^2(t) \rangle = 2 D_l t\,,$ ($l=x,y,z$) increases linearly in
time along each spatial direction as shown for one parameter set 
in Fig.~\ref{msquaredist}. This behavior
is typical for normal diffusion. Parallel to the
$z$-direction one has a undisturbed diffusion and therefore
the mean square displacement and thus 
$D_z$ is much larger than in the modulated $x$- and $y$-direction.
Along these two directions the saddles and the local maxima between neighboring 
potential valleys act as barriers for the
dumbbell motion and therefore
$D_x$ (equal to $D_y$) is smaller than $D_z$.
Moreover, for a dumbbell 
in a solvent the HI between the two beads 
comes into play which in general enhances the diffusion as
 can be seen by the shift between the solid and the dashed lines in Fig~\ref{msquaredist}.

\begin{figure}[ht]
\vspace{-2mm}
  \begin{center}
\includegraphics[width=0.95\columnwidth]{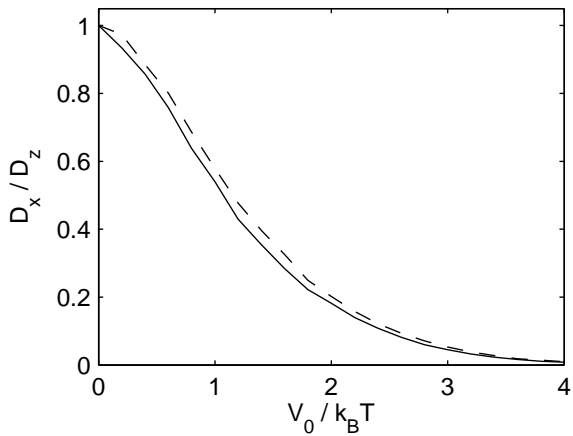}
  \end{center}
\vspace{-5mm}
  \caption{The  diffusion $D_x$ ($=D_y$)
in the $x$-direction is normalized by the free diffusion
$D_z$  in $z$-direction and plotted as a function
of  $V_0/k_BT$ for the case with HI
(solid line) and without HI (dashed line). The following 
parameters have been used: $k=10$ and $\lambda = 2 b$.}
\label{diffpot}
\end{figure}

The decay of the dumbbell diffusion  as a function of the 
ratio between the modulation amplitude of the potential 
and the thermal energy $V_0 / k_BT$ is shown in Fig.~\ref{diffpot} for one parameter set 
with HI (solid line) and without HI (dashed line) between the beads.
The decay of $D_x$ is similar to the results described 
in Refs.~\cite{Lacasta:2004.1,Lacasta:2005.1} on the diffusion of point like particles.
The difference between the two cases with and without HI is shrinking with an increasing  modulation 
amplitude of the potential, because the higher diffusivity caused by HI
becomes less important with increasing potential barriers.  Accordingly,
the ratio $D_x / D_z$ between the diffusion along one modulated direction 
and the unmodulated direction is smaller with HI than for the case
without HI, as shown in Fig.~\ref{diffpot}.

\begin{figure}[ht]
\vspace{-2mm}
  \begin{center}
\includegraphics[width=0.95\columnwidth]{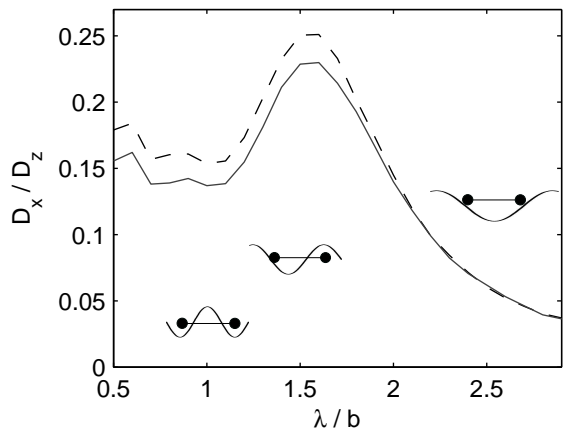}
  \end{center}
\vspace{-5mm}
  \caption{The normalized diffusion constant $D_{x}/D_z$ ($=D_y/D_z$) 
of the dumbbell is shown as a function of the ratio between the equilibrium
 bead distance $b$ and the wavelength $\lambda$ for the case  with HI (solid line) and without HI (dashed line). The insets illustrate
possible locations of the dumbbell with respect to the
periodic potential for different values of $\lambda/b$.
The parameters are $ V_0/k_BT = 2$ and $k=10$.}
\label{lambdisb}
\end{figure}

In contrast to the diffusion of single point particles the diffusion of a
dumbbell along one modulated direction also depends on 
the interplay between the two  length scales, namely
the bead distance $b$ and the wavelength $\lambda$ of the periodic
potential modulation. A typical functional dependence of the dumbbell
diffusion on the ratio $\lambda/b$ is shown in 
Fig.~\ref{lambdisb}, where the diffusion is remarkably enhanced for $\lambda$ close to $\lambda_1 =3b/2$. Further beyond this value the 
dumbbell diffusion decreases with increasing values of the wavelength up
to a minimum $D_x(\lambda_2)$, which is at about $\lambda_2 \approx 6b$ for the given parameters (not shown in Fig.~\ref{lambdisb}).
The decay of $D_x$ in the range $\lambda_1<\lambda<\lambda_2$ can be explained in the following way.
For increasing  values  $\lambda \gtrsim \lambda_1$ the beads
become essentially caged within one single potential valley 
and an escape from such a trap gets more and more unlikely (see right inset in Fig.~\ref{lambdisb}). If the wavelength 
$\lambda$ is increased further beyond $\lambda_2$ the mean square displacement
and the dumbbell diffusion 
increase again and finally approach
the value of the unmodulated case, because for large wavelengths the dumbbell does not feel the potential anymore.

If on the other hand the modulation wavelength is reduced below $\lambda_2$,
the distance from the potential minimum to a saddle 
is shortened. In such a case it 
becomes more likely that the dumbbell is kicked over a saddle to a
neighboring potential valley.
So the diffusion  $D_x$ increases  
with decreasing values in the range $\lambda_1 < \lambda < \lambda_2$.

Near the  maximum of the diffusion constant at $\lambda \approx \lambda_1 $ 
an additional effect comes into play which supports a
dumbbell movement from one potential valley to another and 
accordingly enhances the dumbbell diffusion. 
In this range the two beads hardly fit into one single potential valley as indicated by the middle inset in Fig.~\ref{lambdisb}. Moreover, for a rather
stiff spring both beads can not reach 
the minima of two neighboring potential valleys simultaneously. So the required excitation energy is
smaller than $V_0$ and for this reason the diffusive motion is enhanced.

In the range $\lambda \leq b$ the dumbbell easily finds an orientation in the potential plane where the two beads are located in two distinct minima, cf. the left inset of Fig.~\ref{lambdisb}. Only one bead needs to be kicked to another valley in order to make progress for the CM of the dumbbell. Therefore, irrespective whether the two beads belong to next nearest neighbor valleys or not, the required excitation energy for a shift of the CM depends only weakly on $\lambda$. This is the origin of the small variations of $D_x$ in the range $\lambda < b$. 

\begin{figure}[ht]
\vspace{-2mm}
  \begin{center}
\includegraphics[width=0.95\columnwidth,height=170pt]{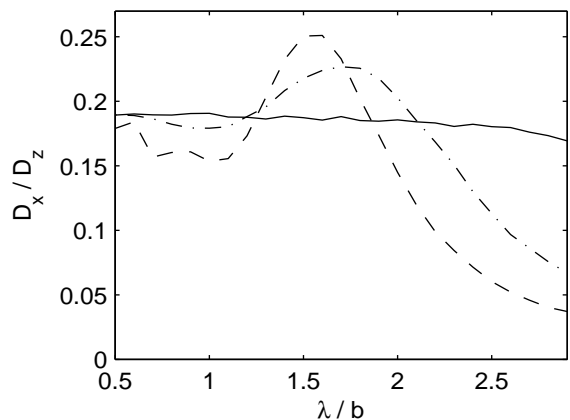}
  \end{center}
\vspace{-5mm}
  \caption{The ratio between the diffusion constants $D_x/D_z$ 
of the dumbbell  is shown for three different values of the 
spring constant (solid line: $k=0.1$, dashed-dotted: $k=1$, dashed: $k=10$) as a function of the ratio $\lambda/b$. The potential amplitude is $ V_0 = 2 k_BT$.}
\label{klambdisb}
\end{figure}

\begin{figure}[ht]
  \begin{center}
 \includegraphics[width=0.95\columnwidth,height=170pt]{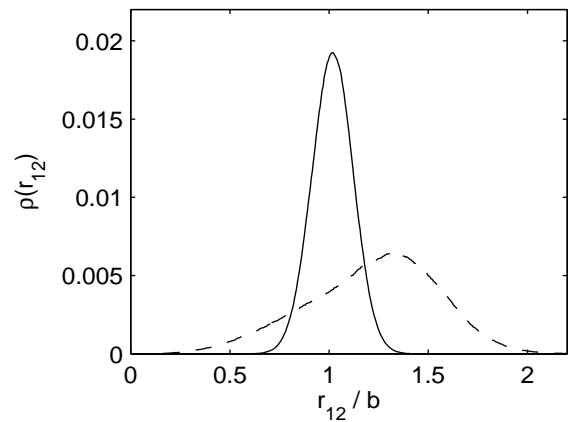}
  \end{center}
\vspace{-5mm}
  \caption{The distribution $\rho( r_{12} )$ of the bead distances is shown
for two different values of the spring constant,  $k=10$ (solid line) and $k=1$ (dashed line) and for the parameters  $V_0/k_BT=2$ and $\lambda/b = 3/2$.}
\label{distribution}
\end{figure}

The explanation given for the local maximum of the diffusion constant in the range of $\lambda_1$ in
Fig.~\ref{lambdisb} is supported by the influence of the spring constant $k$ on the height of $D_x(\lambda_1)$ and on
the mean distance $\langle r_{12} \rangle$.
The local maximum of $D_x$ is especially pronounced in the case of
a rather stiff dumbbell (see Fig.~\ref{klambdisb}) where the distribution of the bead
distance $\rho(r_{12})$ is not changed by the periodic potential (see Fig.~\ref{distribution}).
On the other hand for smaller values of $k$
the maximum of $\rho(r_{12})$ is more and more shifted from $b$ to $\lambda$.
In this case the beads relax down to the potential valleys, so a higher excitation energy is required for a shift of the
CM, which results in a smaller diffusion constant as indicated  by the
solid line in  Fig.~\ref{klambdisb}.

\vspace{-2mm}
\begin{figure}[ht]
  \begin{center}
 \includegraphics[width=0.95\columnwidth,height=220pt]{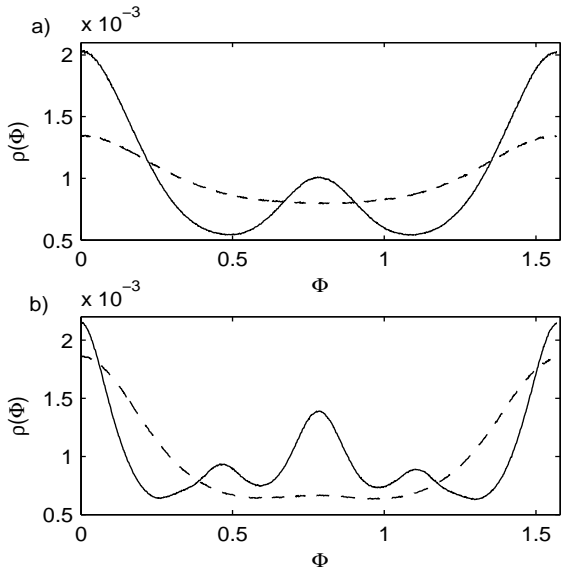}
  \end{center}
\vspace{-5mm}
  \caption{The orientational distribution $\rho(\Phi)$ of the  dumbbell 
axis in the $xy$-plane is shown for two different values of the spring 
constant (solid line: $k=1$, dashed: $k=10$) and for $V_0/k_BT=2$.
The angle $\Phi$ is measured with respect to the $x$-axis and the ratio between the wavelength $\lambda$ and the bead distance $b$ is
in part  a) $\lambda/b=2$ and in part b) $\lambda/b=1$.}
\label{angledist}
\end{figure}
\vspace{-2mm}

The distance between the potential valleys depends
on the direction in the $xy$-plane and therefore
the orientational distribution of the dumbbell axis $\rho(\Phi)$ in Fig.~\ref{angledist}, 
provides a complementary  information to  $\rho( r_{12} )$.
The two beads of the dumbbell may relax more easily 
to the potential minima in the case of a soft spring (cf. solid lines in Fig.~\ref{angledist}) compared to a stiff spring (cf. dashed lines).
So the orientational distribution of the dumbbell axis in the 
$xy$-plane shows, besides the maxima along the $x$- and the $y$-direction, a local maximum for a diagonal 
orientation of the dumbbell axis. This is displayed in part a) of Fig.~\ref{angledist}. If the wavelength is reduced
to $\lambda=b$ one finds, in addition to the maximum in 
the  diagonal direction (11), local maxima along  
the (21)- and the (12)-direction, 
as indicated by part b) of Fig.~\ref{angledist}.

{\it Conclusions.-}\label{sec: conclusions} 
The Brownian motion of a dumbbell in a two-dimensional periodic 
potential has been investigated in terms of a Langevin model. 
For an increasing barrier height 
between the potential minima, we find a decreasing diffusion constant for the center of mass
along the modulated spatial directions as well as a reduction of the influence of the hydrodynamic interaction.
For stiff springs the diffusion additionally depends on the ratio between
the wavelength $\lambda$ of the potential modulation 
and the equilibrium dumbbell extension $b$. In the range $\lambda \approx 3b/2$ this interplay is especially pronounced, 
because the two beads do not fit into one single or two neighboring potential minima anymore and this mismatch causes a reduced effective barrier height and thus an enhanced diffusion constant. If the spring constant is small the beads can relax down to the potential minima over a wide range of $\lambda$, which results in a diffusion constant that is rather independent from $\lambda$ in this domain. So the height of the maximum of the diffusion constant at $\lambda \approx 3b/2$ increases with the spring stiffness.
For modulation wavelengths further beyond $3b/2$
the diffusion constant decays monotonically 
until some minimum is reached. In this
range the dumbbell is essentially caged in one single potential
valley and it is rather unlikely that it escapes.
Beyond this minimum as $\lambda$ goes to infinity
the diffusion constant grows until the free diffusion limit is reached.

The presented results may be useful for sorting polydisperse particle mixtures with 
respect to the particles' elasticity and size.

{\it {Acknowledgments.-}}
We would like to thank L. Holzer for instructive discussions.
This work has been supported by the German science foundation through
the priority program on micro and nanofluidics SPP 1164.
\vspace{-3mm}

\bibliographystyle{prsty}

\end{document}